\documentclass[a4paper]{jpconf}
\usepackage{graphicx}
\usepackage{mathptmx}

\begin{document}
\title{PandaX-III: Searching for Neutrinoless Double Beta Decay with High Pressure Gaseous Time Projection Chambers}

\author{Ke Han for the PandaX-III Collaboration}

\address{INPAC and School of Physics and Astronomy, Shanghai Jiao Tong University;
Shanghai Laboratory for Particle Physics and Cosmology, Shanghai 200240, China}

\ead{ke.han@sjtu.edu.cn}

\begin{abstract}
The PandaX-III (Particle And Astrophysical Xenon Experiment III) experiment will search for Neutrinoless Double Beta Decay (NLDBD) of $^{136}$Xe at the China Jin-Ping underground Laboratory II (CJPL-II).
In the first phase of the experiment, a high pressure gas Time Projection Chamber (TPC) will contain 200 kg, 90\% $^{136}$Xe enriched gas operated at 10 bar.
Microbulk Micromegas, a fine pitch micro-pattern gas detector, will be used for charge readout and enable us to reconstruct tracks of NLDBD events with good energy and spatial resolution.
With simulation, we demonstrate excellent background suppression capability with tracking information.
In this proceeding, we will give an overview of recent progress of PandaX-III, including data taking of a 20~kg scale prototype TPC.
\end{abstract}

\section{Introduction}

Neutrinoless double beta decay (NLDBD) is one of the most sought-after topics in particle physics nowadays.
The discovery of such a process would prove the Majorana nature of neutrinos, demonstrate lepton number violation and thus have far reaching implications beyond neutrino physics~\cite{Avignone:2007fu, Elliott:2014iha}.
Around the world, there are more than a dozen of experiments searching for NLDBD of various candidate isotopes.
The goal of current and next generation of NLDBD experiments is to cover the phase space of inverted neutrino mass hierarchy, or about 15~meV in terms of effective Majorana mass.
The nominal target isotope mass is around 1~ton to reach such a goal.
For a recent review of different experimental status, one can find in~\cite{Ostrovskiy:2016uyx} and in the proceedings of this conference.
The PandaX-III experiment~\cite{Chen:2016qcd} aims to search for NLDBD of $^{136}$Xe with high pressure  gaseous Time Projection Chambers (TPCs).
The first phase will feature one 200~kg TPC module and the second phase will consist of five upgraded modules for a ton-scale experiment.
The detectors will be hosted under an ultra-clean water shielding at China Jin-Ping underground Laboratory II (CJPL-II), where cosmic rays and other environmental background can be effectively suppressed.

The PandaX-III gaseous TPC can image tracks of emitted electrons from NLDBD and thus provide powerful background suppression and signal selection capability.
%The hypothetical NLDBD events feature two emitted electrons carrying a sum of energy of the decay Q-value on the MeV scale.
%The sum of energies measured experimentally is only broadened by instrumentation resolution.
A common goal of NLDBD experiments is to reconstruct a spectrum of total energy of two electrons and to identify a possible signal peak over background around Q-value.
Almost all the experiment, with the notable exceptions such as NEXT~\cite{Lopez-March:2017qzh} and NEMO-3/SuperNEMO~\cite{Waters:2017wzp}, measure only energies of electrons.
PandaX-III, built on the successful T-REX R\&D effort~\cite{Irastorza:2015dcb}, is expected to have a good energy resolution of 3\% Full-Width-Half-Maximum (FWHM) at the Q-Value of 2.458~MeV.
%Two electrons emitted from possible NLDBD will go through multiple scattering in the high pressure gas medium and eventually lose all the energy.
More importantly, PandaX-III can image the track of NLDBD electrons, which is typically on the order of 10~cm in 10~bar xenon, with millimeter level granularity.
As energy of each electron decreases along its travel, energy loss per unit length increases dramatically, which is called Bragg peak in the $dE/dx$-$x$ plot.
Therefore a signature of NLDBD events would be one bright blob due to Bragg-peaking at each end of the track.
However, for background events such as gamma rays, only one bright blob is present.
The distinctive feature of NLDBD events can be fully explored with the so-called \emph{topological analysis} with PandaX-III gaseous TPC.
%Later we will show that the background suppression power can be a factor of about 40 while maintaining reasonably high signal efficiency.

The PandaX-III collaboration has recently published the conceptual design of the first 200~kg scale TPC~\cite{Chen:2016qcd} and built a 20~kg scale prototype. In this proceeding, we will firstly describe the design features and physics reach of the full-size module. The construction, commissioning, and first results from the prototype TPC will also be discussed.
\section{PandaX-III TPC with Microbulk Micromegas}

\begin{figure}[tb]
\begin{center}

\begin{minipage}{0.4\textwidth}
\includegraphics[width=\textwidth]{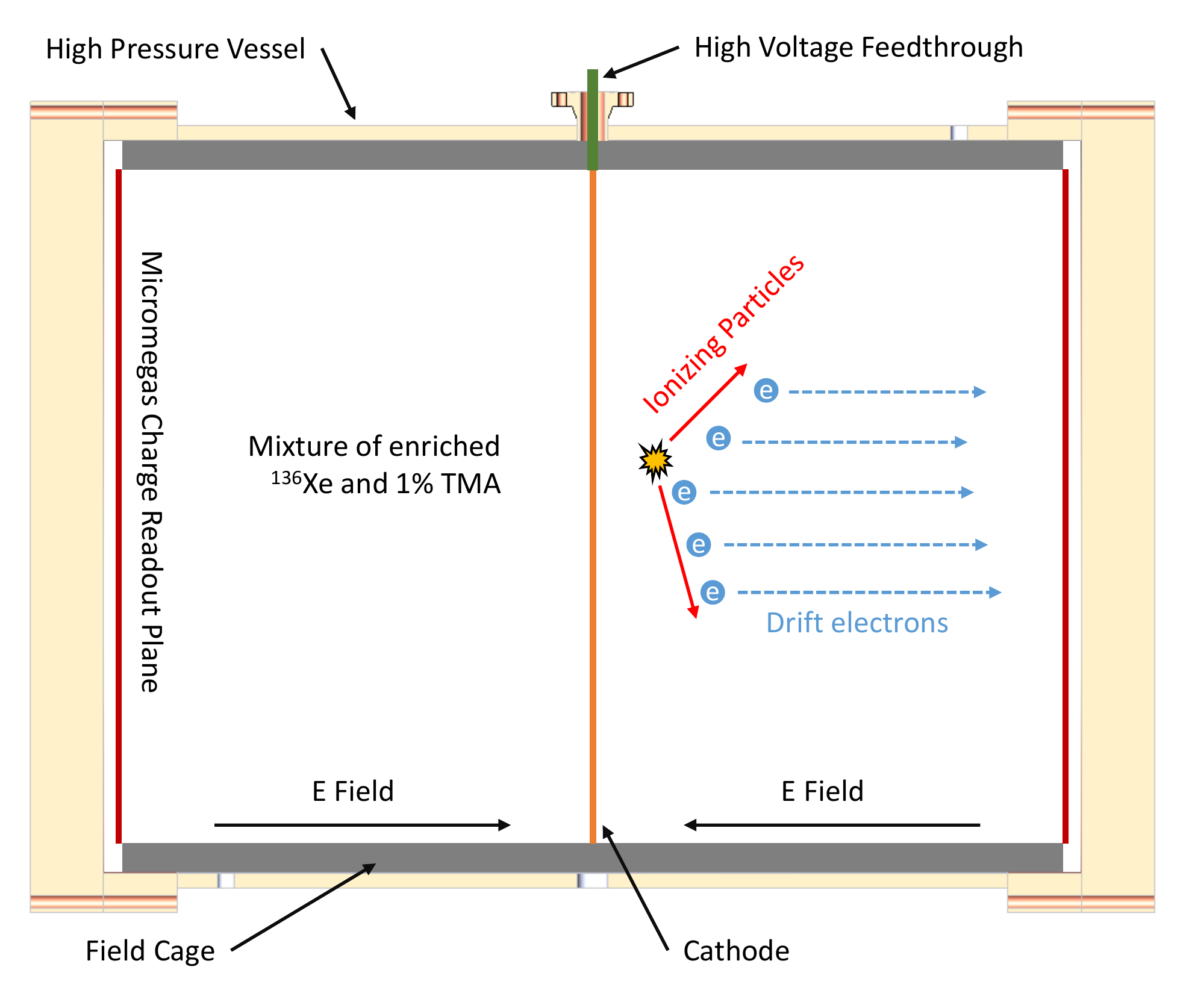}
\end{minipage}
\begin{minipage}{0.5\textwidth}
\includegraphics[width=\textwidth]{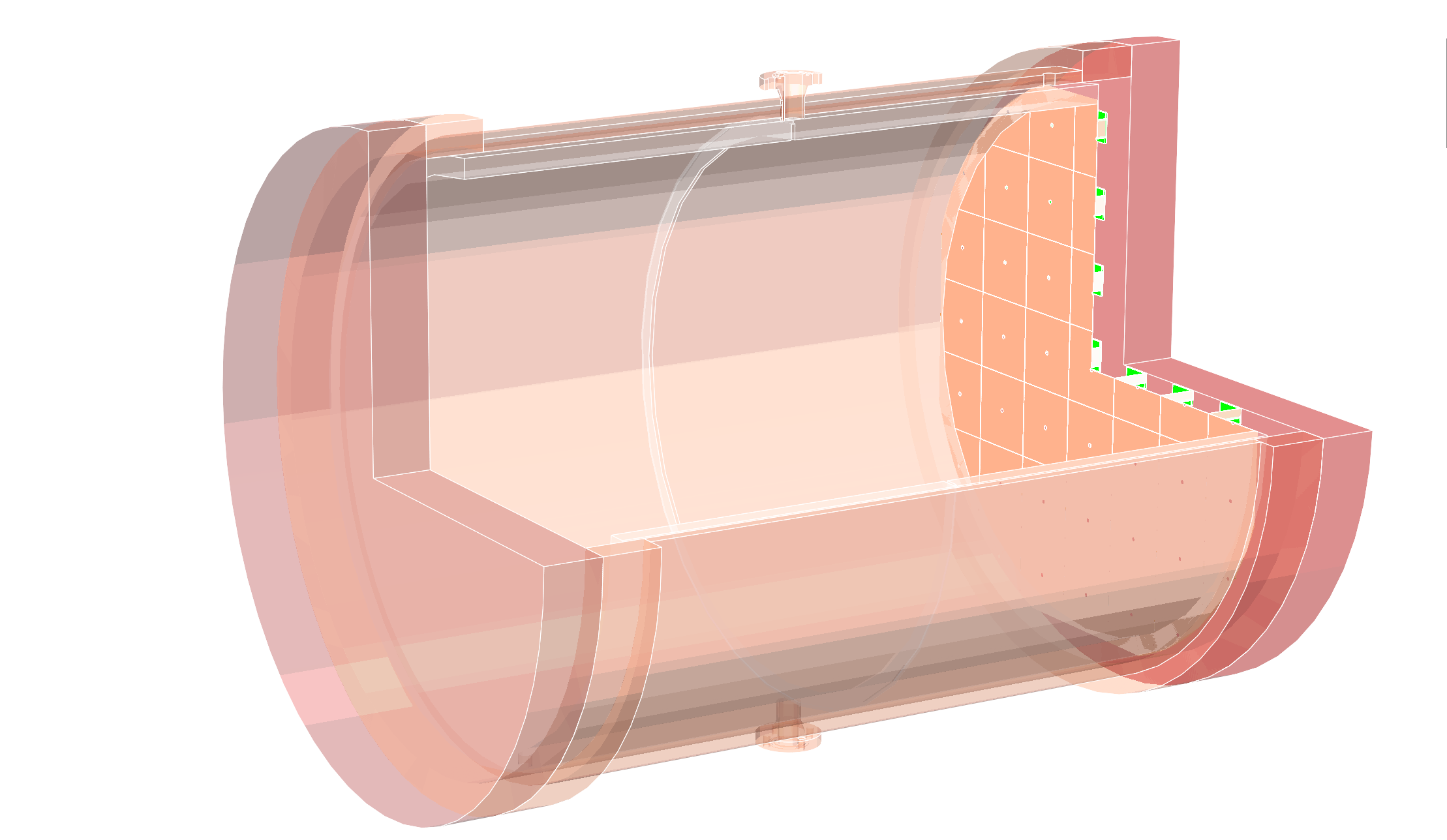}
\end{minipage}
\end{center}
\vspace{-1em}
\caption{ (Left) Illustration of the PandaX-III TPC with main features highlighted. (Right) A cut-away of the TPC.}
\label{fig:TPC}
\end{figure}

The first PandaX-III module will have a cylindrical active volume of 1.5~m in diameter and 2~m in length.
The TPC will have a symmetrical design with a cathode in the middle and a meter-long drift field for each half (Figure~\ref{fig:TPC}).
Negative High Voltage (HV) to the cathode will be provided through a HV feedthrough from the side of the pressure vessel.
The TPC will hold about 200~kg of enriched $^{136}$Xe gas at 10~bar, which serves as the source of NLDBD as well as the TPC medium.
%Any event such as a NLDBD signal would ionize xenon gas and ionized electrons will drift towards the charge readout plane at the end of the vessel.
A quencher gas, 1\% of TMA (trimethylamine) is mixed with xenon to suppress scintillation of xenon and convert more energy to ionization.
It also reduces electron diffusion while drifting and offers a more stable operation condition.

Event energy and tracks are measured with charge readout planes.
Each charge readout plane consists of 41 20~cm by 20~cm Micromegas modules.
Micromegas is a micro-pattern gaseous detector which amplifies and collects electrons~\cite{Giomataris:1998rc}.
The Micromegas we use is of the Microbulk type and fabricated out of Kapton and copper with standard lithography work-flow~\cite{Andriamonje:2010zz}.
The amplification gap is of 50~$\mu$m long and determined largely by the thickness of Kapton films.
The uniformity of gap distance and thus uniformity of the gain is superior to the traditional bulk type Micromegas.
With the intrinsically radio-pure materials, Microbulk Micromegas (MM) is especially suitable for rare event searches.
MM we use has strip readout, which means pixels along $X$ (or $Y$) directions are grouped together for readout.
The pitch size is 3~mm and there are a total of 128 channels per module.
The 200~kg TPC would have a total over 10000 readout channels.
Custom electronics with commercial AGET chips will be used for reading out such a large number of channels.

\section{Background estimation with topological analysis}

Background estimation is studied with Monte Carlo simulation in three stages.
The first stage is to treat PandaX-III detector as a simple calorimeter and reconstruct energy deposition of signals and background events in the whole detector.
In the second stage, we added detector response, including drift electron transportation and diffusion in the field cage, triggering, and sampling rates, and calculated background budget with those factors considered.
It's also worth noting that event tracks are also generated in this stage as a mock data set.
The simulated tracking information is used in the last stage, the so-called topological analysis, to further select signals and reject background events.
Topological analysis is developed at this moment for analyzing simulated data set but will also be applicable to future detector data.
Finally, the energy spectrum around the Q-value after all the cuts is used for sensitivity projection.

PandaX-III uses two Geant4-based Monte Carlo simulation packages, REST and BambooMC, to generate simulated data set and calculate the background budget.
REST is developed as a general-purpose simulation and analysis framework for gaseous TPCs.
BambooMC has been used widely and successfully in the previous generations of dark matter experiments, including PandaX-II~\cite{Tan:2016zwf}.
Both packages built independent geometry from detector design and gave consistent background budget values.
Here we briefly discuss about the simulation input and background budget.
A lot more details, with tables outlining contributions from each components, can be found in~\cite{Chen:2016qcd}.

\subsection{PandaX-III as a calorimeter}

\begin{figure}[tb]
\begin{center}
\includegraphics[width=0.4\textwidth]{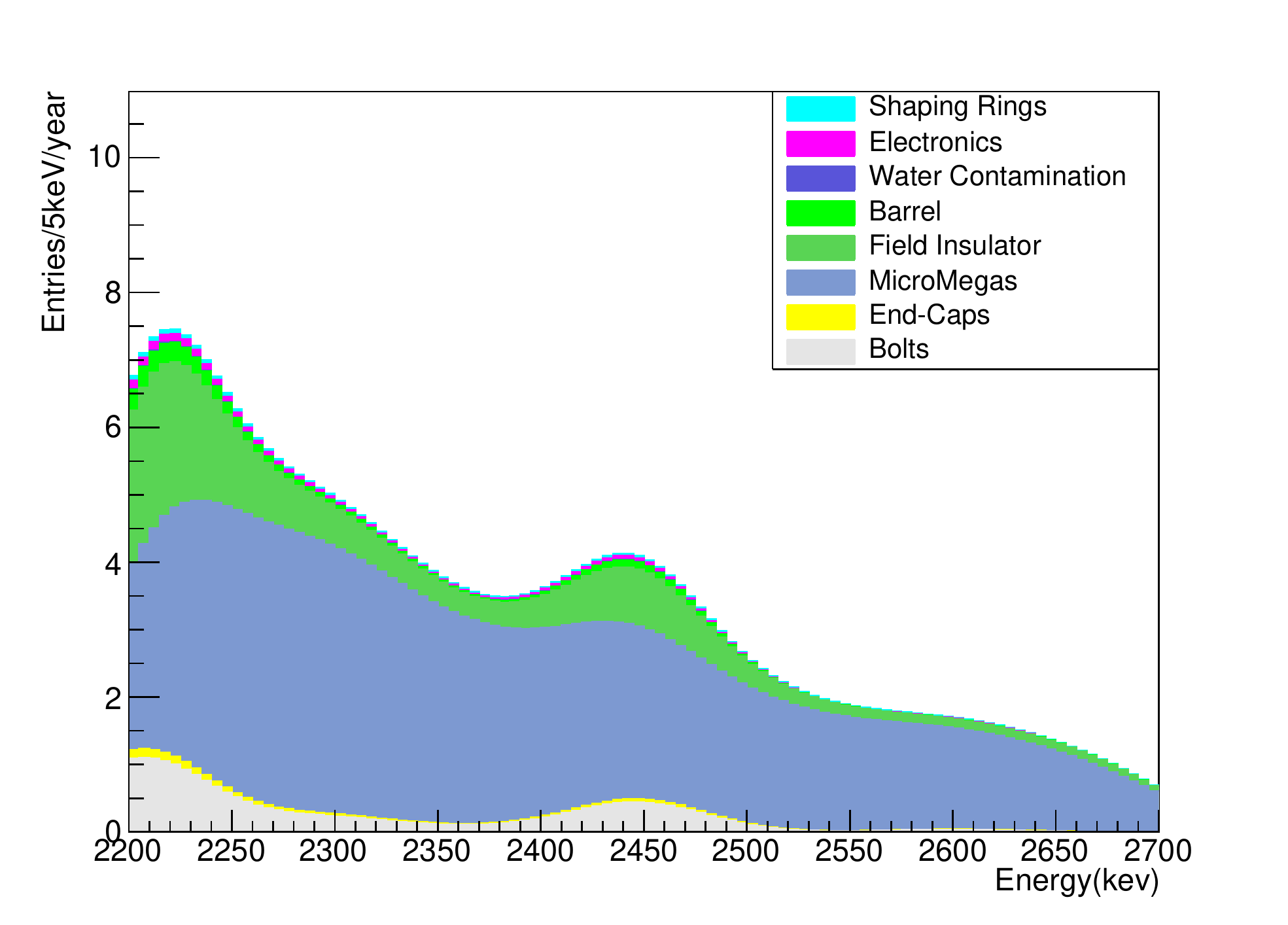}
\hspace{0.05\textwidth}%
\includegraphics[width=0.4\textwidth]{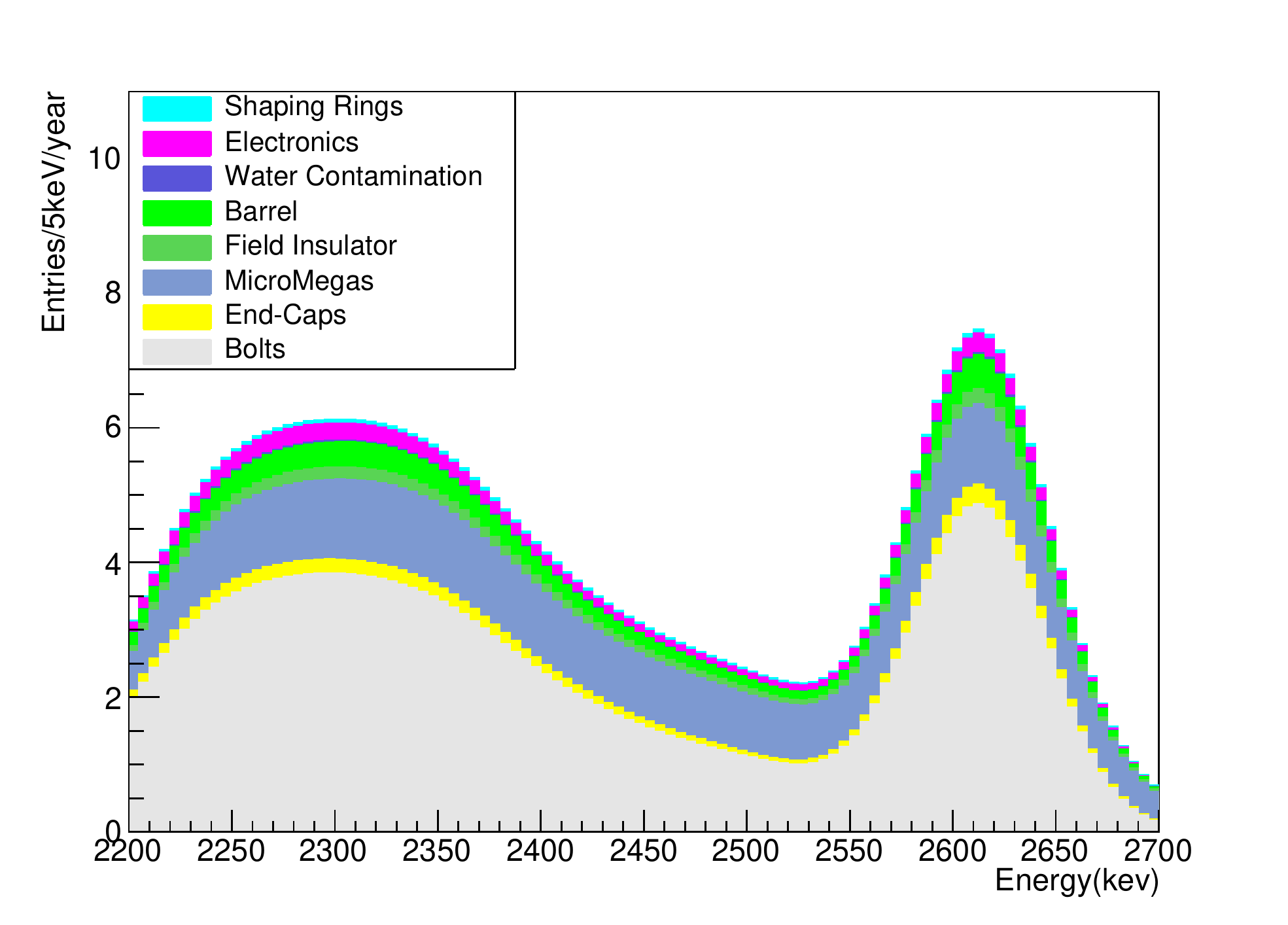}%\hspace{1pc}%
\end{center}
\vspace{-1.5em}
\caption{Stack histograms of simulated background contribution from different laboratory and detector components for $^{238}$U (Left) and $^{232}$Th (Right).
3\% FWHM energy resolution is assumed at the Q-value = 2458 keV.}
\label{fig:RawSpectra}
\end{figure}

We consider background contribution from the underground laboratory environment, passive water shielding, and the detector itself.
More specifically, radioactive background from laboratory walls, ultra-pure water from the water shielding pool, copper pressure vessel (copper end-caps, copper barrel, and stainless steel bolts included), electronics, filed cage, Micromegas charge readout plane, and cathode are simulated.
For each components, other than laboratory wall and ultra-pure water, we used best-available published radioactive contamination measurements (or upper limits) as input for our simulation.
The output energy spectra for contaminations of U and Th chains are shown in Figure~\ref{fig:RawSpectra}, assuming an energy resolution of 3\% FWHM at the Q-value.
The energy smearing is equivalent to detector effect such as Micromegas gain fluctuation, gain non-uniformity, the drift electron loss in the field cage, etc.

\subsection{Detector response from a gaseous TPC}
Additional gaseous TPC specific features, such as time window cuts, are added to the simulation.
The time window, define the time span of drift electrons hitting the charge readout plane, is proportional to the track length in $Z$ direction (drift direction) and electron drifting velocity.
For the majority of NLDBD events in PandaX-III, the track length in $Z$ is around 10~cm and time window is within 102.4~$\mu$s.
However, if bremsstrahlung happens, the track may become discontinuous and span a much larger time (distance) window.
Even though a NLDBD event may deposit the full 2458~keV within the detector active volume, the detector can not collect all the electrons due to limited time window.
A gamma background may have more discontinuous tracks than a NLDBD signal because of the multiple Compton scattering nature and are cut by time window more effectively.

Another detector hardware cut is the active readout area of Micromegas.
On each charge readout plane, 41 Micromegas modules are tiled together and cover most but not all of the 1.5~m diameter active area.
After this cut, the total signal efficiency is calculated to be 54.2\% by BambooMC.

Electron diffusion and strip readout of Micromegas module are also implemented in this stage, which provides the two 2-dimensional tracks ($XZ$ and $YZ$) for topological analysis in the next stage.

\subsection{Topological analysis}

\begin{figure}[tb]
\begin{minipage}{0.6\textwidth}
\includegraphics[width=\textwidth]{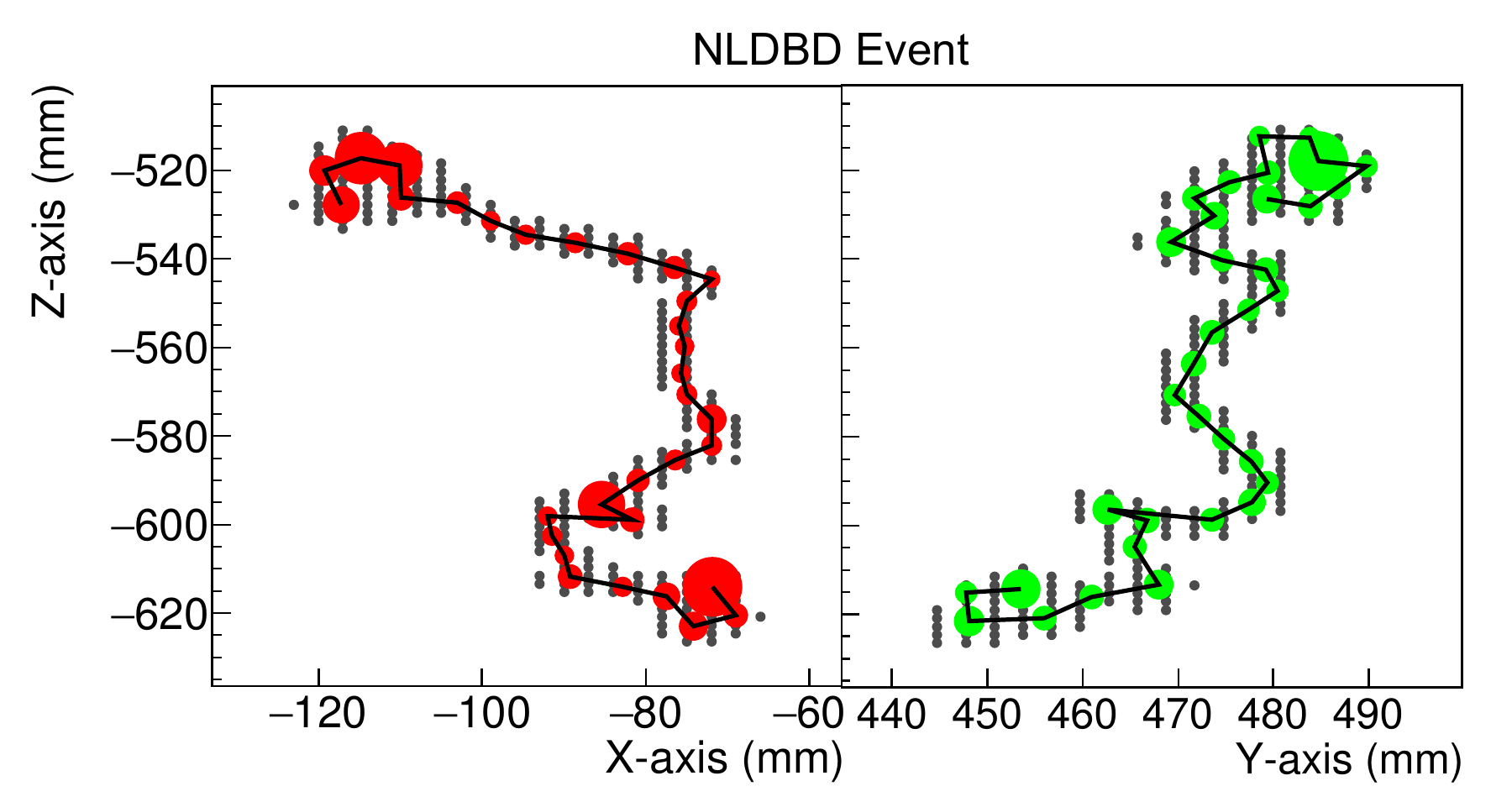}
\end{minipage}\hspace{2pc}%
\begin{minipage}{0.3\textwidth}
\includegraphics[width=\textwidth]{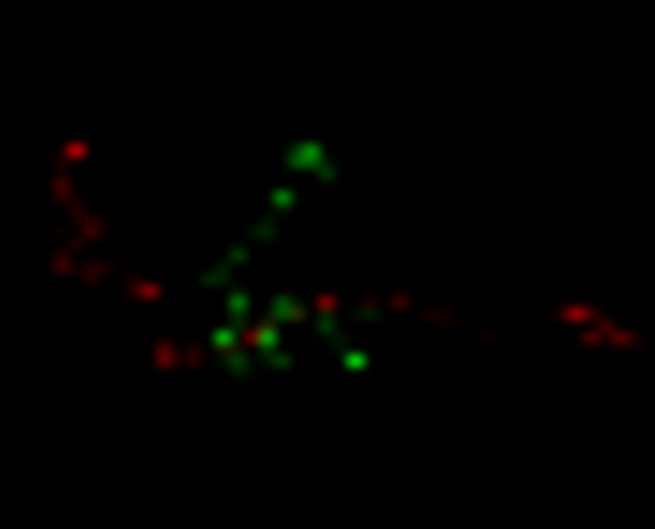}
\end{minipage}
\caption{Illustration of topological analysis with traditional track- and blob-finding algorithms (Left) and an example input image for classification with neural network.}
\label{fig:topology}
\end{figure}

We developed two parallel approaches to identify the Bragg-peaking blobs at both ends of NLDBD events.
The traditional approach finds a track to connect all the discrete pixels (strips) in the $XZ$ and $YZ$ planes, as illustrated in Figure~\ref{fig:topology} (Left).
Then energy deposition per unit area along the track is calculated.
If passing certain threshold value, two large blobs at two ends would indicate a possible NLDBD events.
Preliminary analysis shows that the track-finding and blob-finding algorithms can keep about 59\% of the signals and reject over 97\% of the background events.

Another approach is to unitize the Convolutional Neural Network (CNN) for image pattern recognition.
The $XZ$ and $YZ$ 2D hit map is converted to $red$ and $green$ channels of an input image (see Figure~\ref{fig:topology} (Right)), which is feed into CNN for classification.
MC generated signal and background events are used for training and validation of the CNN.
The best performing CNN would reject 98\% of the background while keeping the signal efficiency as high as 80\%, according to our preliminary studies.

\section{Prototype TPC: commissioning and first results}

\begin{figure}[tb]
\begin{minipage}{0.25\textwidth}
\includegraphics[width=\textwidth]{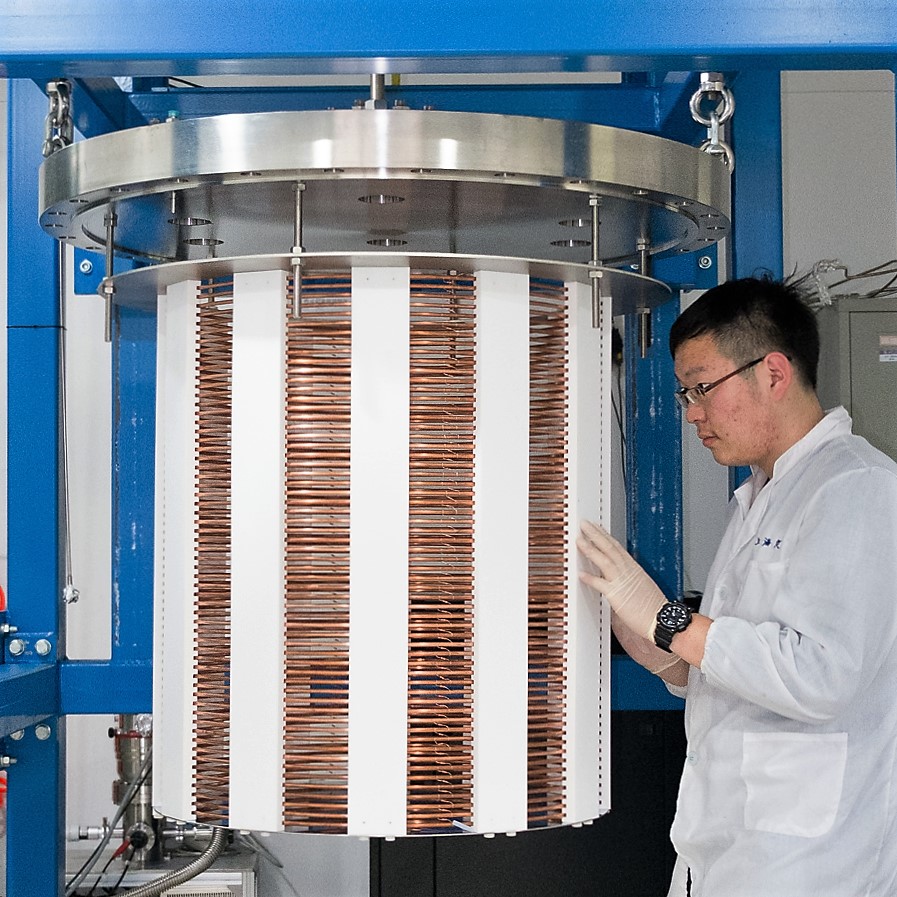}
\end{minipage}\hspace{2pc}%
\begin{minipage}{0.35\textwidth}
\includegraphics[width=\textwidth]{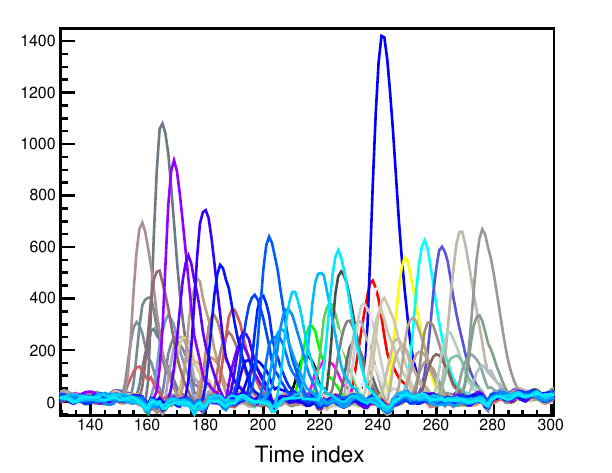}
\end{minipage}
\begin{minipage}{0.35\textwidth}
\includegraphics[width=\textwidth]{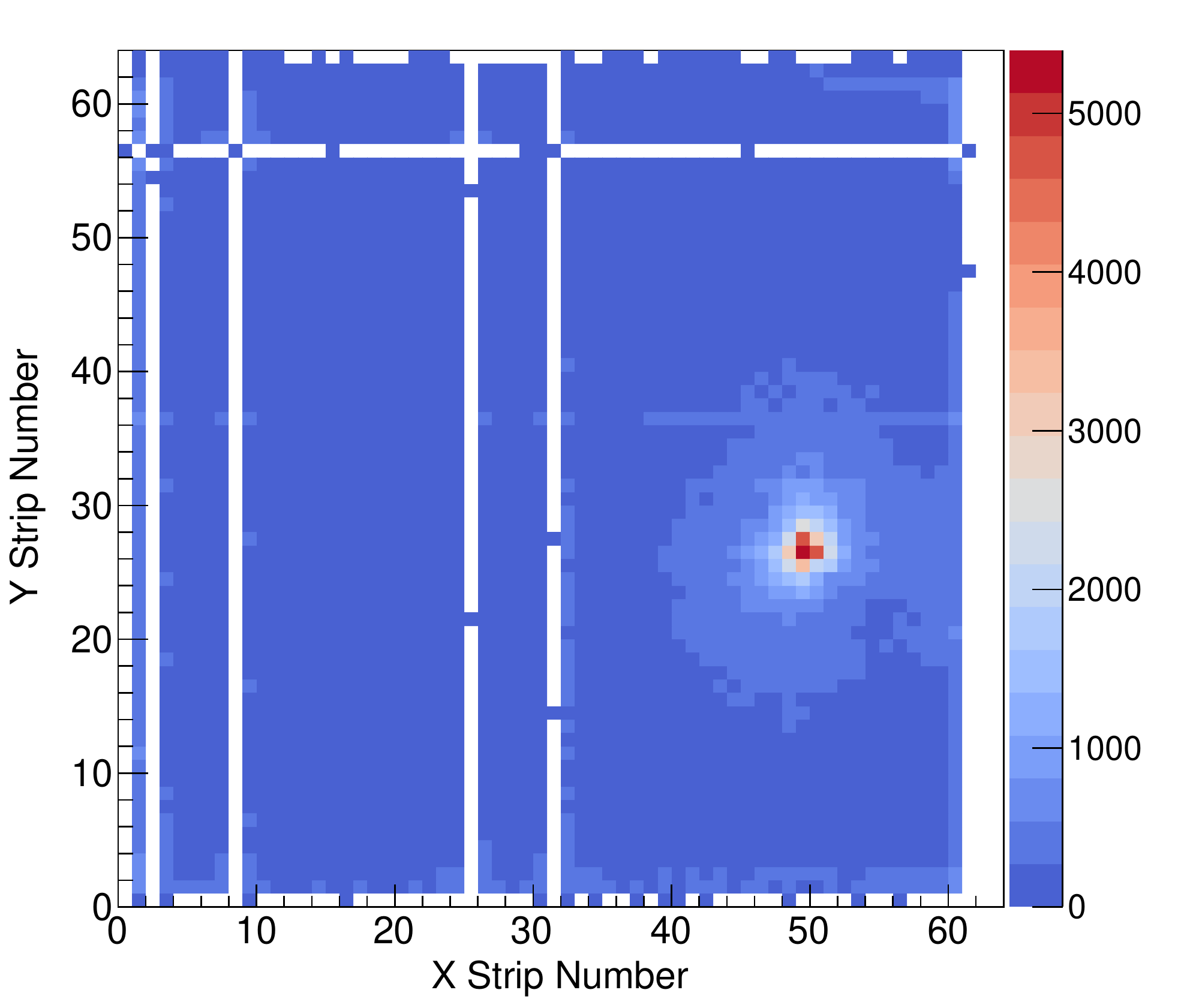}
\end{minipage}
\caption{(Left) A picture of Prototype TPC. (Middle) Example pulses on strips for cosmic muon event. (Right) Image of a $^{241}$Am source in the field cage.}
\label{fig:Prototype}
\end{figure}

To study the performance of large high pressure gaseous TPC, especially MM with strip readout, we have built a 20~kg scale prototype TPC.
The TPC is single-ended, with the cathode at the bottom and charge readout plane on top.
Its cylindrical shape has a diameter of 66~cm and a height of 78~cm.
At 10~bar, it holds about 20~kg of xenon gas in the active volume.
Figure~\ref{fig:Prototype} (Left) shows the field cage with a student to scale.

The charge readout plane can accommodate up to 7 MM modules, 2 of which are installed at this moment.
We have run the TPC with different gas mixture, including pure argon, pure xenon, Ar+(5\%)Isobutene, Ar+(30\%)CO$_2$, Xe+(1\%)TMA, with pressure as high as 5~bar.
The middle panel of Figure~\ref{fig:Prototype} shows a cosmic muon event in Ar+(5\%)Isobutene mixture.
Sources such as $^{241}$Am, $^{55}$Fe, $^{109}$Cd have been used for energy calibration.
$^{241}$Am is especially useful since it has a high energy alpha line as well as multiple gamma lines from 10 to 60 keV, originating from $^{241}$Am and its daughter $^{237}$Np.
The hot spot in Figure~\ref{fig:Prototype} (Right) shows an example image of the $^{241}$Am source hanging about 40~cm below the charge readout plane.

\section{Conclusion and outlook}
\begin{figure}[tb]
\begin{minipage}{0.45\textwidth}
\vspace{-0.8em}
\includegraphics[width=\textwidth]{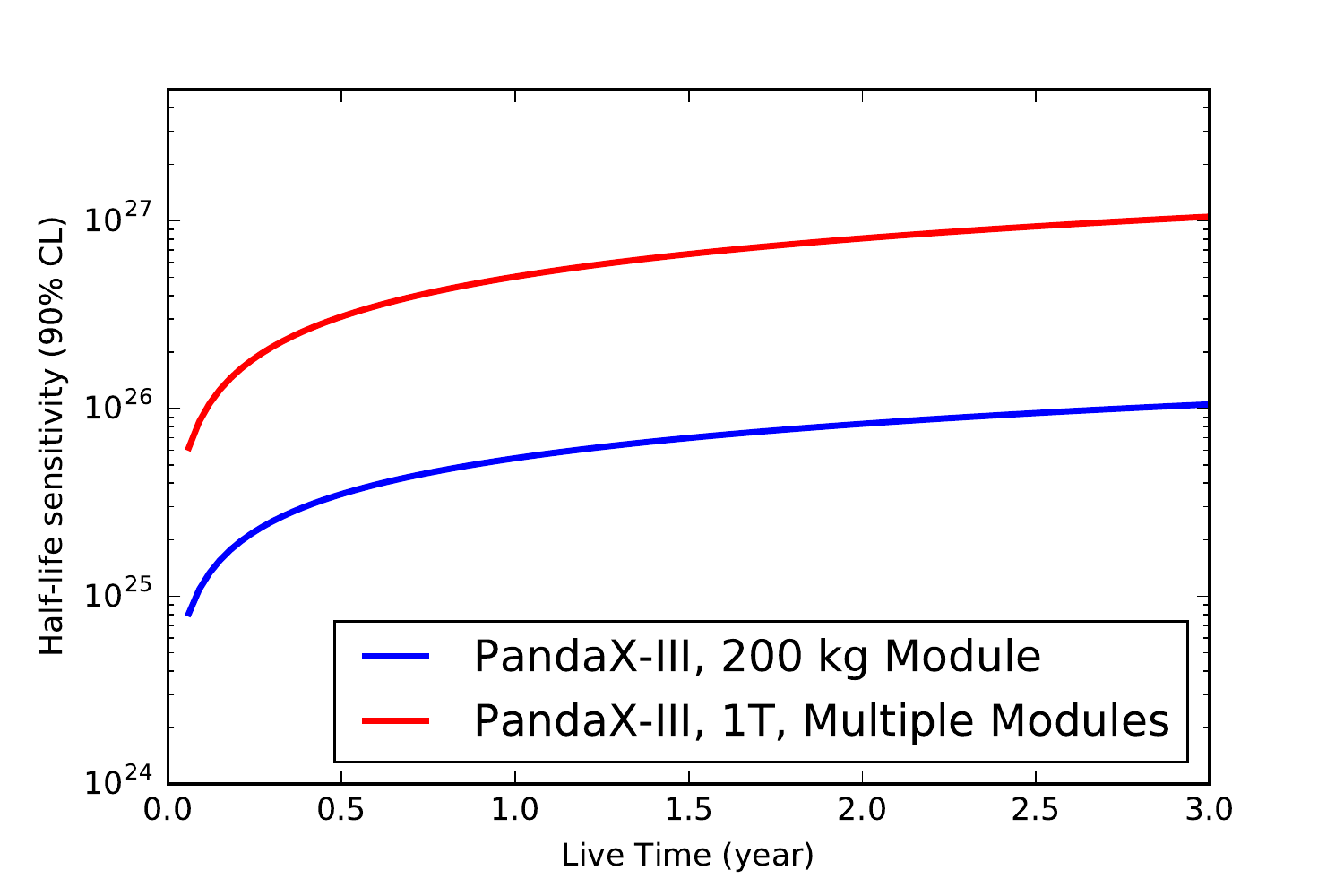}
\end{minipage}
\begin{minipage}{0.55\textwidth}

\includegraphics[width=\textwidth]{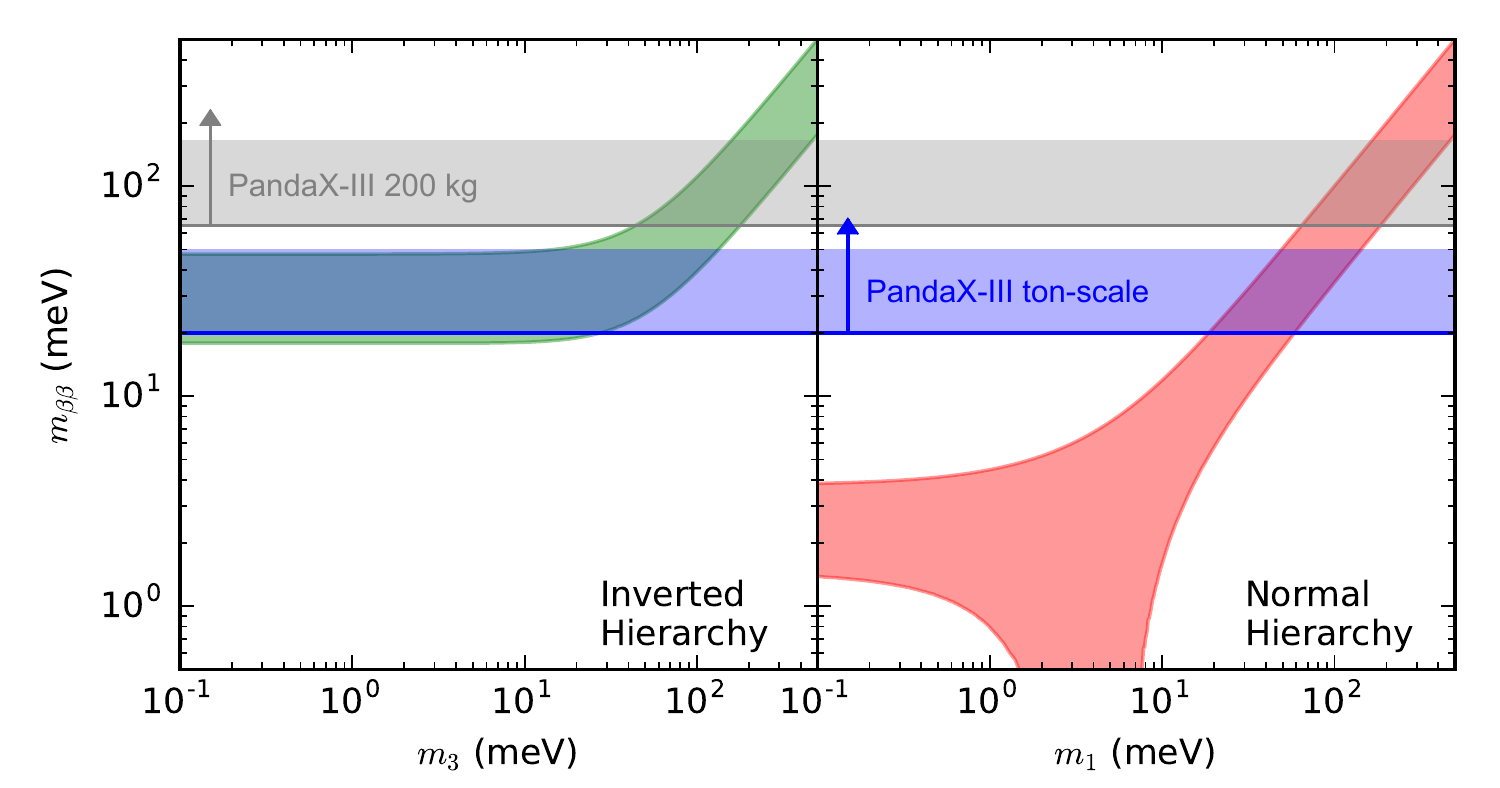}
\end{minipage}
\caption{(Left) PandaX-III sensitivity (90\% CL) to the half-life of $^{136}$Xe NLDBD vs. live time. (Right) Effective Majorana mass sensitivity of PandaX-III with 3-year live time.}
\label{fig:Sensitivity}
\end{figure}

With a projected energy resolution of 3\% FWHM, a signal efficiency of 35\%, and a background rate of $10^{-4}$ c/keV/kg/yr after topological cuts, the first PandaX-III module can reach a half-life sensitivity (90\% CL) of $10^{26}$ years after 3 years of live time (See Figure~\ref{fig:Sensitivity} (Left)).
For the future ton scale experiment, we assume an upgraded energy resolution of 1\% and a lower background rate of $10^{-5}$ c/keV/kg/yr.
The ultimate sensitivity after 3 years would reach $10^{27}$ years.
The corresponding effective Majorana mass $m_{\beta\beta}$ is 65 to 165~meV and 20 to 50~meV for the first module and ton-scale setup respectively (See Figure~\ref{fig:Sensitivity} (Right)).

The PandaX-III collaboration focuses on technical design of subsystems, radio-assay of detector materials, and commissioning of the prototype TPC with 7 Micromegas.
The installation of 7 Micromegas in the prototype TPC is going on at this moment and is expected to to take calibration data by the end of 2017.
With the calibration data, we can study carefully energy and track reconstruction of high energy events, further optimize the design of Micromegas modules, and explore the impact of absolute $Z$ positioning.
Construction of the full TPC is expected to start after summer 2018.

\ack{This work was supported by the National Key Programme for Research and Development (NKPRD) Grant \#2016YFA0400300 from the Ministry of Science and Technology, China. This project has been supported by a 985-III grant from Shanghai Jiao Tong University and the Key Laboratory for Particle Physics, Astrophysics and Cosmology, Ministry of Education, China.}

\section*{References}
\bibliography{PandaX-III_ref}

\end{document}